\documentstyle[11pt,epsf,twocolumn,cuted,twoside]{article}
\begin{document}
\begin{strip}
\vspace*{-2.5cm}
\title{Electroweak phase transition and Higgs self-couplings in the two-Higgs-doublet model}
\author{S. W. Ham$^{(1)}$ and S. K. Oh$^{(1,2)}$
\\
\\
{\it $^{(1)}$ Center for High Energy Physics, Kyungpook National University}\\
{\it Daegu 702-701, Korea} \\
{\it $^{(2)}$ Department of Physics, Konkuk University, Seoul 143-701, Korea}
\\
\\
}
\date{}
\maketitle
\vspace*{-1cm}
We calculate both the cubic and the quartic self-couplings of the lighter scalar Higgs boson
without assuming the decoupling limit in the two-Higgs-doublet model (THDM).
In some regions of parameter space of the THDM where the electroweak phase transition is
strongly first order, it is possible that the quartic self-coupling of the lighter scalar Higgs boson
might be deviated by at least 40 \% from the standard model prediction.
\end{strip}
\section{Introduction}

Explaining the observed baryon asymmetry of the universe is regarded as one of the basic features
for theoretical models to be phenomenologically realistic.
Several decades ago, Sakharov suggested  three essential conditions for dynamically generating
the baryon asymmetry: the violation of baryon number conservation, the violation of both C and CP,
and the deviation from thermal equilibrium [1].
Among various mechanisms for explaining the baryon asymmetry, many attentions have been paid to
the baryogenesis via the electroweak phase transition (EWPT) [2], which in principle may satisfy
the three Sakharov conditions.
As is well known, in order to ensure sufficient deviation from thermal equilibrium, the EWPT should be
first order, and its strength should be strong, since otherwise the baryon asymmetry generated
during the phase transition subsequently would disappear.
In general, the phase transition is regarded as a strongly first order if the critical value of
the vacuum expectation value of the Higgs field at the broken state is
larger than the critical temperature.

The Standard Model (SM), has been investigated whether it can realize the EWPT.
It is found, however, that the EWPT faces severe difficulty to be realized in the SM,
because the strength of the EWPT is too weak in the SM for the present experimental lower
bound on the mass of the Higgs boson.
Thus, in the SM, the EWPT is weakly first order or higher order for the experimentally
allowed mass of the Higgs boson [3].
Also, the Cabibbo-Kobayashi-Maskawa matrix in the SM [4] cannot produce large enough CP violating phase
for generation of baryon asymmetry.
Thus, the idea of baryogenesis via the EWPT requires extensions of the SM to be realized.
Many scenarios have been studied in the literature [5-8].

Among them is the two-Higgs-doublet model (THDM) [9].
The presence of an additional Higgs doublet in the THDM enables either explicit or
spontaneous CP violation to occur in the Higgs sector of the THDM [10].
Also, it has been observed that there are parameter regions in the THDM where the EWPT is
strongly first order to generate the desired baryon asymmetry.
Quite recently, phenomenological implications of the THDM within the context of the EWPT
have been considered by Okada and his colleagues [11].
In particular, they have suggested that the cubic self-coupling of the lighter scalar Higgs boson
in the THDM might be considerably different from the SM prediction.
This information would be very useful and duly tested at the future $e^+e^-$ ILC.
In fact, it has already been addressed that the knowledge of both the cubic and
the quartic self-couplings of the Higgs bosons is essential for the reconstruction
of the necessary self-interaction Higgs potential [12].

We are motivated by the article by Okada and his colleagues, and we would like to study
in more detail the self-couplings of the Higgs bosons.
In particular, we examine if the quartic self-coupling of the lighter scalar Higgs boson
in the THDM might be significantly different from the SM prediction, as well as the cubic self-coupling,
by studying the finite temperature effective Higgs potential in the THDM at the one-loop level,
without decoupling limit, under the condition of the strongly first order EWPT.
We find that not only the cubic self-coupling but also the quartic self-coupling of the Higgs bosons
in the THDM exhibits a large deviation from the SM predictions, in the parameter regions
where strongly first order EWPT is possible, for values of the lighter scalar Higgs boson mass
between 120 and 210 GeV.

\section{The Higgs sector without decoupling limit}

Following the notations of Ref. [11], the most general form of the Higgs potential of the THDM
at the tree level is given in terms of two Higgs doublets, $\Phi_1$ and $\Phi_2$, by
\begin{eqnarray}
    V_{\rm tree} & = &
    m_1^2 |\Phi_1|^2 + m_2^2 |\Phi_2|^2
    - (m_3^2 \Phi_1^{\dagger} \Phi_2 + {\rm H.c.} ) \cr
    & &\mbox{}
    + {\lambda_1 \over 2} |\Phi_1|^4 + {\lambda_2 \over 2} |\Phi_2|^4
    + \lambda_3 |\Phi_1|^2 |\Phi_2|^2 \cr
    & &\mbox{}
    + \lambda_4 |\Phi_1^{\dagger} \Phi_2|^2
    + \left [{\lambda_5 \over 2} (\Phi_1^{\dagger} \Phi_2)^2 + {\rm H.c.} \right ] ,
\end{eqnarray}
where $\lambda_i$ are quartic couplings and $m_i$ $(i=1,2,3)$ are the mass parameters.
The discrete $Z_2$ symmetry is softly broken by the term proportional to $m^2_3$,
which prevents the flavor changing neutral current process in the tree level.
Assuming no CP violation in the Higgs sector of the THDM, we have five physical Higgs bosons
with definite CP after electroweak symmetry breaking: Two neutral scalar Higgs bosons ($h, H$),
one neutral pseudoscalar Higgs boson ($A$), and a pair of charged Higgs bosons ($H^{\pm}$).
It is understood that $h$ is lighter than $H$.

We take $m_1 = m_2 = m$ and $\lambda_1 = \lambda_2 = \lambda'$ [11],
which reduces the field direction relevant to the electroweak phase transition
to $\langle \Phi_1\rangle = \langle \Phi_2 \rangle = (0, \varphi / 2)$,
which corresponds to $\sin (\alpha - \beta) = -1$ and $\tan \beta = 1$,
where $\tan \beta$ is the ratio of the vacuum expectation values $v_2$
of $\Phi^0_2$ to $v_1$ of $\Phi^0_1$ and $\alpha$ is the mixing angle between the two scalar Higgs bosons.
The vacuum expectation values satisfy $v = \sqrt{2 (v^2_1 + v^2_2)} = 246$ GeV.
In terms of $\varphi$, the tree-level Higgs potential is given by
\begin{equation}
V_0 (\varphi) = - {\mu^2 \over 2} \varphi^2 + {\lambda \over 4} \varphi^4 ,
\end{equation}
where $\mu^2 = m_3^2 - m^2$ and $\lambda = (\lambda' + \lambda_3 + \lambda_4 + \lambda_5)/4$.
The tree-level masses of Higgs bosons, $h$, $H$, $A$, and $H^{\pm}$, are given as
$m_h^2 = 2 \lambda v^2$, and $m_{\phi}^2 = M^2 + \lambda_{\phi} v^2$
$(\phi = A, H, H^{\pm})$, where $M^2 = 2 m_3^2 /\sin 2 \beta$ and $\lambda_{\phi}$
are the linear combinations of $\lambda_1$-$\lambda_5$.
In the decoupling limit, where $M^2$ is very larger than $v^2$, $h$ behaves as the SM Higgs boson
and the masses of the heavier Higgs bosons are dominantly dependent on $M$.
We are interested in the non-decoupling limit, where $M^2$ is not so large, $h$ might
behave differently from the SM Higgs boson, and the masses of the heavier Higgs bosons are
at most a few hundred GeV.
Nevertheless, we set $m_{\phi} = m_A = m_H = m_{H^{\pm}}$ for the heavy Higgs boson masses,
for simplicity.

The one-loop effective potential at zero temperature $V_1 (\varphi, 0)$ is obtained
from the effective potential method as [13]
\begin{eqnarray}
    V_1(\varphi, 0) & = &
    \sum_l {n_l \over 64 \pi^2}
        \left [ m_l^4 (\varphi) \log \left ({m_l^2 (\varphi) \over m_l^2 (v) } \right ) \right. \cr
        & &\mbox{}
    \left. - {3 \over 2} m_l^4 (\varphi) + 2 m_l^2 (v) m_l^2 (\varphi) \right ] ,
\end{eqnarray}
where $l$ stands for various participating particles: the gauge bosons $W$, $Z$,
the third generation quarks $t$, $b$, and the Higgs bosons $h$, $H$, $A$, $H^{\pm}$.
The degrees of freedom for each particle are:
$n_W = 6$, $n_Z = 3$, $n_t = n_b = - 12$, $n_h = n_H = n_A = 1$, and $n_{H^{\pm}} = 2$.

The finite-temperature contribution at the one-loop level to the Higgs potential is given by [14]
\begin{eqnarray}
    V_1 (\varphi, T) & = &
    \sum_{l=B,F} {n_l T^4 \over 2 \pi^2}
    \int_0^{\infty} dx \ x^2 \ \log \left [1 \right. \cr
    & &\mbox{}
    \left. \pm \exp{\left ( - \sqrt {x^2 + {m_l^2 (\varphi)/T^2 }} \right )  } \right ] ,
\end{eqnarray}
where the negative sign is for bosons ($B$) and the positive sign for fermions ($F$).

One may employ the high temperature approximation to obtain an analytical expression of
$V_1(\varphi, T)$ for qualitative discussions on electroweak phase transition.
It is known that in the SM the high temperature approximation is consistent
with the exact calculation of the integrals within 5 \% at temperature $T$ for $m_F/T < 1.6$
and $m_B/T < 2.2$, where $m_F$ and $m_B$ are the mass of the relevant fermion and boson, respectively.
Explicitly, in the high temperature approximation, $V_1(\varphi, T)$ may be expressed as
\begin{equation}
     V_1(\varphi, T) \simeq (D T^2 - E) \varphi^2 - F T \varphi^3 + G \varphi^4  ,
\end{equation}
where
\begin{eqnarray}
    D & = & {1 \over 24 v^2} (\sum_B n_B m_B^2 + 6 m_t^2 + 6 m_b^2 ) , \cr
    E & = & {m_h^2 \over 4} - {1 \over 32 \pi^2 v^2} ( \sum_{l=B,F} n_l m_l^4) , \cr
    F & = & {1 \over 12 \pi v^3} (\sum_B n_B m_B^3) , \cr
    G & = & {m_h^2 \over 8 v^2}
    - {1 \over 64 \pi^2 v^4} \left [\sum_{l=B,F} n_l \log {m_l^2 \over a_l T^2} \right ] ,
\end{eqnarray}
with $\log (a_F) = 1.14$ and $\log (a_B) = 3.91$.
Also we note that the above one-loop effective potential contains the lighter scalar Higgs boson
contribution.
As one can see from the above expressions, the first order electroweak phase transition is
strengthened by the term proportional to $F$ due to the heavier Higgs boson contributions.
If the contributions of heavier Higgs bosons be negligible, $V_1(\varphi, T)$ would reduce
to contain the contribution of only $h$ in the Higgs sector, thus would become approximately
equivalent to the SM.
In this case, the electroweak phase transition is either weakly first order or higher order.
We perform the exact calculation of the integrals in  $V_1(\varphi, T)$ instead of employing
the high temperature approximation.

The full effective potential at finite temperature at one-loop level may now be expressed as
\[
    V(\varphi, T) = V_0(\varphi,0) + V_1(\varphi, 0) + V_1(\varphi, T)  .
\]
We emphasize that the above one-loop effective potential contains the contribution of
the lighter scalar Higgs boson.

At the one-loop level, the cubic and the quartic self-couplings of the scalar Higgs boson
in the SM are respectively given by
\begin{eqnarray}
    \lambda_{hhh}^{\rm SM} & \simeq &
    {3 m_h^2 \over v} \left [1 + \sum_l {n_l m_l^4 \over 12 \pi^2 v^2 m_h^2} \right ] , \cr
    \lambda_{hhhh}^{\rm SM} & \simeq &
    {3 m_h^2 \over v^2} \left [1 + \sum_l {4 n_l m_l^4 \over 12 \pi^2 v^2 m_h^2} \right ] ,
\end{eqnarray}
where $l$ stands for the gauge bosons, the third generation quarks, and the SM Higgs boson.
On the other hand, the cubic and the quartic self-couplings of the lighter scalar Higgs boson
in the THDM at the one-loop level are respectively given by
\begin{eqnarray}
    \lambda_{hhh}^{\rm THDM} & \simeq &
    \lambda_{hhh}^{\rm SM}
    + {m_{\phi}^4 \over \pi^2 v^3} \left (1 - {M^2 \over m_{\phi}^2} \right )^3 , \cr
    \lambda_{hhhh}^{\rm THDM} & \simeq &
    \lambda_{hhhh}^{\rm SM}
    + {m_{\phi}^4 \over 2 \pi^2 v^4} \left (1 - {M^2 \over m_{\phi}^2} \right )^3 \cr
    & &\mbox{}
    \times \left (2 + {M^2 \over m_{\phi}^2} \right ) ,
\end{eqnarray}
where the contributions of heavier Higgs bosons ($\phi = H, A, H^{\pm}$) can be collected
into one term since $m_{\phi} = m_H = m_A = m_{H^{\pm}}$.

\section{Numerical Analysis}

Now, we define the deviations of the cubic and the quartic self-couplings of the lighter scalar
Higgs boson in the THDM from those in the SM Higgs boson, respectively, as
\begin{eqnarray}
    \Delta_{hhh} &=&
    (\lambda_{hhh}^{\rm THDM}-\lambda_{hhh}^{\rm SM})
    /\lambda_{hhh}^{\rm SM}  , \cr
    \Delta_{hhhh} &=&
    (\lambda_{hhhh}^{\rm THDM}-\lambda_{hhhh}^{\rm SM})
    /\lambda_{hhhh}^{\rm SM} .
\end{eqnarray}
For numerical analysis, we set $m_W$ = 80.425 GeV, $m_Z$ = 91.187 GeV, $m_t$ = 174.3 GeV,
and $m_b$ = 4.2 GeV.
The remaining free parameters are $m_h$, $M$, and $m_{\phi}$.

For $m_h = 120$ GeV, we examine $V(\varphi, T)$ for $0  \le M \le 160$ GeV and
$150 \le m_{\phi} \le 400$ GeV, by adjusting the temperature $T$ to the critical temperature $T_c$.
We find that $T_c = 120.7$ GeV for $m_h = 120$ GeV.
We establish the contour of $v_c/T_c = 1$ in the ($M$, $m_{\phi}$)-plane, where $v_c$ is
defined as the distance between the two degenerate vacua at $T_c$.
Our result is shown in Fig. 1 as the solid curve of Set 1.
Above the solid curve (the region of larger $m_{\phi}$ values) is the region
where a strongly first-order EWPT is allowed.
We also calculate $\Delta_{hhh}$ and $\Delta_{hhhh}$ for the same $m_h = 120$ GeV,
and plot the result in Fig. 1.
The contour of $\Delta_{hhh} = 6$ \%, and that of $\Delta_{hhhh} = 43$ \% are shown
respectively as the dashed curve and the dotted curve of Set 1.
As one can see, both contours lie below the contour of $v_c/T_c = 1$ for the whole parameter space
in the ($M$, $m_{\phi}$)-plane.
This implies that it is possible for some region of parameter space of the THDM where the EWPT
is strongly first-order that either $\Delta_{hhh} \ge 6$ \% and/or $\Delta_{hhhh} \ge 43$ \%.
In other words, the THDM may allow a significant deviation from the SM when the EWPT is
strongly first-order.
Our study of $\Delta_{hhh}$ for $m_h = 120$ GeV is consistent with the result of Ref. [11].
On the other hand, the result of $\Delta_{hhhh}$ is new.

Now, we repeat the analysis for some different values of  $m_h$.
Our results are shown in Fig. 1 as Set 2, 3, and 4.
The relevant numbers of Set 2 are:  $m_h = 150$ GeV, $T_c = 134.8$ GeV, $\Delta_{hhh} = 7.5$ \%,
and $\Delta_{hhhh} = 44$ \%;
those of Set 3 are: $m_h = 180$ GeV, $T_c = 147.4$ GeV, $\Delta_{hhh} = 10$ \%,
and $\Delta_{hhhh} = 55$ \%;
and those of Set 4 are: $m_h = 210$ GeV, $T_c = 159.0$ GeV, $\Delta_{hhh} = 13$ \%,
and $\Delta_{hhhh} = 62$ \%.

\setcounter{figure}{0}
\def\figurename{}{}%
\renewcommand\thefigure{Fig. 1}
\begin{figure}[t]
\epsfxsize=7.5cm
\hspace*{5mm}
\epsffile{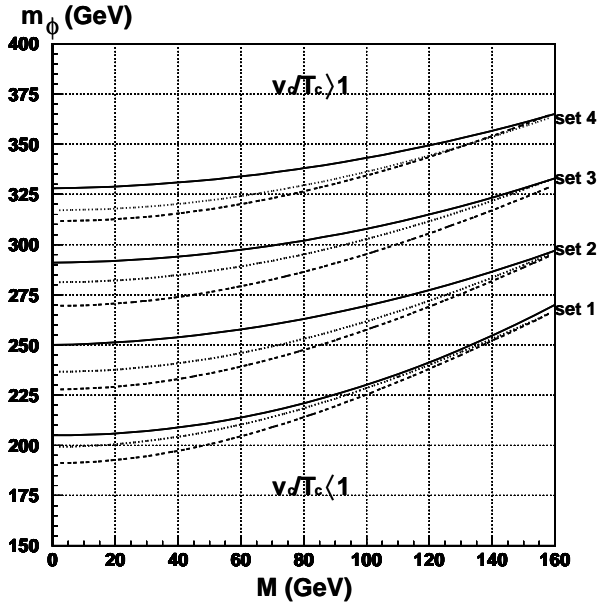}
\vspace*{-10mm}
\caption[plot]{Contours of $v_c/T_c = 1$ (solid curve), $\Delta_{hhh}$ (dashed curve),
 and $\Delta_{hhhh}$ (dotted curve), for $m_h = 120$ GeV (Set 1), $m_h = 150$ GeV (Set 2),
 $m_h = 180$ GeV (Set 3), and $m_h = 210$ GeV (Set 4) in the $(M, m_{\phi})$-plane.
 The values of $\Delta_{hhh}$ and $\Delta_{hhh}$ in each Set are different. See the text.}
\end{figure}

\section{Conclusions}

We find that the results of our study are consistent with the suggestions made by Okada and
his colleagues.
They have found that the THDM allows a strongly first-order EWPT for successful baryogenesis
of the universe for some parameter region, where the cubic self-coupling of the lighter scalar
Higgs boson might be significantly affected.
Motivated by the article by Okada and his colleagues, we have studied not only
the cubic self-coupling but also the quartic self-coupling of the lighter scalar Higgs boson in the THDM.

We have explored a wide range of value of the lighter scalar Higgs boson mass.
Our calculations have been done under the assumption of no CP violation, non-decoupling limit,
with reasonable simplifications in relevant parameter values.
The results are quite remarkable, especially in the case of the quartic self-coupling.
If the mass of the lighter scalar Higgs boson is 210 GeV, it is possible
that the quartic self-coupling of the lighter scalar Higgs boson in the THDM may be larger than
that of the SM by more than 60 \% for some parameter values where the EWPT is strongly first-order.
It is found that the cubic self-coupling of the lighter scalar Higgs boson does not exhibit
its deviation from the SM as vivid as the quartic self-coupling does.
The cubic self-coupling of the lighter scalar Higgs boson in the THDM may be larger than
that of the SM by about 13 \% if the mass of the lighter scalar Higgs boson is 210 GeV.
For smaller mass of the lighter scalar Higgs boson, the possible magnitude of the deviation
in the quartic self-coupling becomes small, but never negligible.
We suggest that the deviation of the self-couplings in the THDM from that of the SM, which is
induced by the non-decoupling effects of the loops of heavier Higgs bosons at the one-loop level,
might provide some basis for the THDM to be further investigated at the future
International Linear Collider, ILC.

\vskip 0.3 in

\noindent
{\large {\bf Acknowledgments}}
\vskip 0.2 in
\noindent
This research is supported through the Science Research Center Program
by the Korea Science and Engineering Foundation and the Ministry of Science and Technology.

\vskip 0.2 in



\end{document}